\documentclass[a4paper]{article}

\usepackage[utf8]{inputenc}
\usepackage{graphicx}
\usepackage{a4wide}
\usepackage{url}

\usepackage{enumerate}
\usepackage{amssymb}
\usepackage{amsmath}
\usepackage{amsfonts}
\usepackage{bbm}
\usepackage{latexsym}

\newcommand{\FT}[1]{\widehat{#1}}

\newcommand{\Weql}{W^{(\text{eq})}}

\newcommand{\N}{{\mathbb N}}
\newcommand{\Z}{{\mathbb Z}}
\newcommand{\R}{{\mathbb R}}
\newcommand{\C}{{\mathbb C\hspace{0.05 ex}}}
\newcommand{\vep}{\varepsilon}

\newcommand{\T}{{\mathbb T}}

\newcommand{\rme}{{\rm e}}
\newcommand{\ci}{{\rm i}}
\newcommand{\rmd}{{\rm d}}
\newcommand{\braket}[2]{\langle #1 , #2\rangle}

\newcommand{\defem}[1]{{\em #1\/}} 
\newcommand{\defset}[2]{ \left\{ #1\left|\, #2
\makebox[0cm]{$\displaystyle\phantom{#1}$}\right.\!\right\} }
\newcommand{\set}[1]{\{#1\}}
\newcommand{\norm}[1]{\Vert #1\Vert}

\newcommand{\tr}{\operatorname{Tr}}
\newcommand{\re}{\operatorname{Re}}

\newcommand{\cf}[1]{{\mathbbm 1}_{\{#1\}}}

\newcommand{\CFBN}{\mathcal{C}_{\text{fBN}}}
\newcommand{\Heff}{H^{\text{eff}}}
\newcommand{\CHubb}{\mathcal{C}_{\text{Hubb}}}

\newcounter{jlisti}

\begin{document}

\title{Kinetic theory and thermalization of weakly interacting fermions}

\author{Jani Lukkarinen\thanks{\emailjani}\\[1em] \UHaddress}
\date{\today}
\newcommand{\email}[1]{E-mail: \tt #1}
\newcommand{\emailjani}{\email{jani.lukkarinen@helsinki.fi}}
\newcommand{\UHaddress}{\em University of Helsinki, Department of Mathematics and Statistics\\
\em P.O. Box 68, FI-00014 Helsingin yliopisto, Finland}

\maketitle

\abstract{Weakly interacting quantum fluids allow for a natural kinetic theory description which takes into account the fermionic or bosonic nature of the interacting particles. In the simplest cases, one arrives at the Boltzmann--Nordheim equations for the reduced density matrix of the fluid. We discuss here two related topics: the kinetic theory of the fermionic Hubbard model, in which conservation of total spin results in an additional Vlasov type term in the Boltzmann equation, and the relation between kinetic theory and thermalization.}


\section{Introduction}

Kinetic theory describes motion which is \defem{transport dominated} in the sense that typically the solutions to the kinetic equations correspond to constant velocity, i.e., \defem{ballistic}, motion intercepted by 
\defem{collisions} whose frequency is order one on the kinetic space-time scales.
Weakly interacting quantum fluids provide one such example system, as discussed in detail in \cite{ls09}.  

We focus here on one particular
case of a weakly interacting quantum fluid, the case of weakly interacting fermions hopping on a lattice.
Such a model would arise physically as a description of a fluid of electrons in a crystal background potential.
For our purposes, this model has also other attractive properties, namely, 
it has interesting non-trivial kinetic theory with relatively few technical and mathematical 
difficulties.  Much of the discussion below can be straightforwardly adapted to bosonic lattice systems, at least for initial data which exclude 
formation of Bose--Einstein condensate.  For more details about such extensions, we refer to \cite{ls09}; for instance, Remark 2.3 
summarizes the changes and new properties which arise for fermions and bosons moving not on a lattice, but in the continuum $\R^3$.

The purpose of this contribution is not to provide a comprehensive review of literature on kinetic theory and properties of fermionic systems.  Instead, we focus on building a bridge between mathematically rigorous results and the physics of fermionic systems.  To this end, we begin with a fairly detailed Section 2 on the definition of finite system of fermions hopping on a periodic lattice with pair interactions between the particles, from the point of view of both a fixed particle number Hilbert space and the full antisymmetric Fock space.  In Section 3 we recall how the probabilistic concepts of 
classical particle systems can be generalised into systems of fermions, namely, the definition of quasifree states, reduced density matrices, and truncated correlation functions.  The 
weak coupling limit of the first reduced density matrix of a translation invariant system and its approximation by the appropriate
spatially homogeneous Boltzmann equation is reviewed in Section 4.  As a conclusion, we discuss in Sections 5 and 6 the implications of the properties of the solutions to the Boltzmann equation on thermalization in the original fermionic lattice system.  Acknowledgements and references
can be found at the end of the text.

\section{Dynamics of lattice fermions}

We recall in this section the mathematical description of fermions, possibly with spin,
which are hopping on a finite periodic lattice of length $L \gg 1$. 
The particles move on a lattice whose points are labelled by 
$\Lambda := \Z^d/(L\Z^d)$ which we parametrize by a square centred at the origin.  
For instance, if $L$ is even, we use the parametrisation
\[
 \Lambda = \left\{-\frac{L}{2}+1,\ldots,\frac{L}{2}-1,\frac{L}{2}\right\}^d\,.
\]
In particular, all arithmetic on $\Lambda$ is performed ``modulo $L$'': if $x,y$ are in the above parametrisation of $\Lambda\subset \Z^d$,
then $x+y\in \Z^d$ needs to be identified with its counterpart in the parametrisation.  Explicitly, ``$x+y$'' is equal to 
$x+y-L m\in \Lambda$ where $m\in \Z^d$ is the unique vector for which $x+y-L m\in \Lambda$.

To describe Fourier transforms, we employ the corresponding discrete dual lattice $\Lambda^* := \Lambda/L = (L^{-1} \Z^d)/\Z^d$.  If needed, we use the 
parametrisation implied by the above notation; for instance for even $L$, we use
$\Lambda^* = \left\{-\frac{1}{2}+\frac{1}{L},\ldots,\frac{1}{2}-\frac{1}{L},\frac{1}{2}\right\}^d$.  The arithmetic on $\Lambda^*$ 
is then performed modulo 1, i.e., using the arithmetic inherited from the $d$-torus $\T^d = \R^d/\Z^d\supset \Lambda^* $.

Such periodic arithmetic is particularly well adapted for use of discrete Fourier transforms.
For a function $f:\Lambda \to \C$
we take its Fourier transform to be the function $\FT{f}:\Lambda^* \to \C$ defined by the formula
\[
 \FT{f}(k) := \sum_{x\in \Lambda} \rme^{-\ci 2\pi k\cdot x} f(x)\,,\qquad k\in \Lambda^*\,.
\]
The inverse transform of $g:\Lambda^* \to \C$ is then given by $\tilde{g}:\Lambda \to \C$ defined by 
\[
 \tilde{g}(x) := \int_{\Lambda^*} \!\rmd k\, \rme^{\ci 2\pi k\cdot x} g(k) =
 \frac{1}{|\Lambda|}\sum_{k\in \Lambda^*} \rme^{\ci 2\pi k\cdot x} g(k)\,,\qquad x\in \Lambda\,.
\]
Here and in the following we use the shorthand notation
\[
  \int_{\Lambda^*} \!\rmd k\, \cdots =
 \frac{1}{|\Lambda|}\sum_{k\in \Lambda^*} \cdots \, .
\]
On a finite lattice, the discrete Fourier transform is always pointwise invertible, i.e.,
for all $x\in\Lambda$, $k\in \Lambda^*$,
$(\FT{f})\tilde{\ }(x) = f(x)$, $(\tilde{g})\FT{\ }(k) = g(k)$.

We assume that the dominant free evolution is defined by giving 
the dispersion relation  $\omega:\T^d\to \R$ corresponding to free evolution after a thermodynamic limit $L\to \infty$
has been taken.   More precisely, we let the periodic lattice hopping potential $\alpha:\Lambda \to \R$
be defined by the inverse Fourier transform of the map $\omega|_{\Lambda^*}$, 
\begin{align}\label{eq:defalphaL}
 \alpha(x;L) := \int_{\Lambda^*} \!\rmd k\, \rme^{\ci 2\pi k\cdot x} \omega(k) \,,\qquad x\in \Lambda\,. 
\end{align}
The function $\alpha$ determines the free $n$-particle Hamiltonian $H_0^{(n)}$
by its action on $n$-particle wave vectors $\psi:\Lambda^n \to \C$,
\[
 H^{(n)}_0 \psi(x_1,\ldots,x_n)=\sum^n_{j=1}
\sum_{y\in \Lambda} \alpha(x_j-y)\psi(x_1,\ldots,y,\ldots,x_n) \,.
\]
The above construction allows an $L$-independent diagonalisation of $H^{(n)}_0$ by taking the discrete Fourier transform:
\[
 (H^{(n)}_0 \psi)\FT{\ }(k_1,\ldots,k_n) = \sum^n_{j=1} \omega(k_j) \,\FT{\psi}(k_1,\ldots,k_n)\,.
\]

We assume that the dispersion relation is \defem{smooth} and \defem{symmetric}, $\omega(-k)=\omega(k)$.
Then $\alpha(x;L)$ is always real, and denoting the inverse Fourier transform (i.e., the Fourier series) of $\omega$ by $\alpha$, we then have
$\alpha(x;L)\to \alpha(x)$ for each fixed $x\in\Z^d$ as $L\to \infty$.  In addition, the range of $\alpha$ is finite, 
in the sense that $|\alpha(x)|$ decreases faster than any power as $|x|\to \infty$.

An explicit often considered example case is nearest neighbour hopping.  This corresponds to 
\[
 \omega(k) = c- \sum_{\nu =1}^d \cos(2\pi k_\nu)\,,
\]
where $c\in \R$ is any constant.  For instance, choosing $c=d$, one obtains the standard discrete Laplacian,
\[
 \sum_{y\in \Lambda} \alpha(x-y)\psi(y) = \frac{1}{2}\sum_{\nu =1}^d \left(2 \psi(x) - \psi(x-e_\nu) - \psi(x+e_\nu) \right)\,,
\]
where $e_\nu$ denotes the unit vector in direction $\nu$.  For vectors $\psi$ which are obtained by taking values of a slowly varying function 
$\psi:\R^d\to \C$, 
the right hand side can be approximated by $-\frac{1}{2} \nabla^2 \psi(x)$.  Therefore, in this case one may also think of $H^{(n)}_0$
as a discrete approximation of the standard free $n$-particle Hamiltonian, with particle mass normalized to one.

We construct a pair-interaction potential $V(x;L)$ analogously, starting from its Fourier-transform $\FT{V}:\T^d\to \C$ and defining
\begin{align}\label{eq:defVL}
 V(x;L) := \int_{\Lambda^*} \!\rmd k\, \rme^{\ci 2\pi k\cdot x} \FT{V}(k) \,,\qquad x\in \Lambda\,. 
\end{align}
To make the potential real-valued and symmetric, we assume that $\FT{V}$ is real-valued and symmetric.
The $n$-particle pair-interaction potential $V^{(n)}$ is then defined via the formula
\[
 V^{(n)}(x_1,\ldots,x_n;L) := \frac{1}{2}\sum^n_{i',i=1; i'\ne i} V(x_{i'}-x_i;L)  \,. 
\]
The potential function acts as an multiplication operator on wave-vectors, and we do not make any distinction in the notation
between the function and the operator.  Thus, if $\psi:\Lambda^n \to \C$ is an 
$n$-particle wave vector, then 
\[
 V^{(n)} \psi(x_1,\ldots,x_n)= V^{(n)}(x_1,\ldots,x_n;L) 
 \psi(x_1,\ldots,x_n) \,.
\]
Naturally, if $n=1$, we have $V^{(n)}=0$.

After these preliminaries, we define the full $n$-particle Hamiltonian by choosing an interaction strength $\lambda \ge 0$, setting 
$H^{(0)}_\lambda =0$, and for $n\ge 1$ defining
\[
 H^{(n)}_\lambda := H^{(n)}_0 + \lambda  V^{(n)}\,.
\]
The corresponding evolution equation for $n$-particle wave vectors $\psi(t)$ is 
\[
 \partial_t \psi(t) = -\ci H^{(n)}_\lambda \psi(t)\,.
\]
The $n$-particle Hilbert space is here finite-dimensional, $\mathcal{H}_n = (\C^\Lambda)^{\otimes n}=\C^{\Lambda^n}$. 
By construction, the evolution preserves particle number and each $H^{(n)}_\lambda$ is a bounded self-adjoint operator on 
$\mathcal{H}_n$.  Thus their direct sum $H_\lambda:= \displaystyle\bigoplus_{n=0}^\infty H^{(n)}_\lambda$
defines a self-adjoint operator on  the full Fock space
$\mathcal{F} := \displaystyle\bigoplus_{n=0}^\infty \mathcal{H}_n $. More precisely, the domain of the operator is
\[
 D(H_\lambda) := \defset{\Psi\in \mathcal{F}}{ \sum_{n=0}^\infty \norm{H^{(n)}_\lambda\Psi_n}^2<\infty}\,,
\]
and the action of $H_\lambda$ on $\Psi=(\Psi_0,\Psi_1,\ldots)\in D(H_\lambda)$ yields the vector
$(H^{(n)}_\lambda \Psi_n)_{n=0}^\infty \in  \mathcal{F}$.  (The proof of these properties can be found for instance in \cite[Theorem 2.23]{teschl:quantum}.)  An analogous construction holds for the potential terms $V^{(n)}$ alone, 
and the corresponding full Fock space operator is denoted by $V$; clearly, $\lambda V=H_\lambda-H_0$ on the domain of $H_\lambda$.

Each $H^{(n)}_0$ and $V^{(n)}$ clearly commutes with permutations of particle labels (i.e., with all of the operators $Q_\pi$
defined by $(Q_\pi\psi)(x_1,\ldots,x_n) = \psi(x_{\pi(1)}, \ldots,x_{\pi(n)})$, there $\pi$ is any permutation of $\set{1,2,\ldots,n}$).
Thus $H_\lambda$ leaves invariant both the fermionic Fock space $\mathcal{F}_-$, containing those $\Psi\in \mathcal{F}$ for which each $\Psi_n$ is antisymmetric under permutations of particle labels, and the bosonic  Fock space $\mathcal{F}_+$, containing only symmetric $\Psi_n$.

From now on, we focus on the corresponding \defem{fermionic} lattice system which is defined by wave vectors $\Psi(t)\in \mathcal{F}_-$
and the semigroup generated by the restriction of $H_\lambda$ to $\mathcal{F}_-$.  Since wave vectors with only finitely many non-zero particle sectors
belong to $D(H_\lambda)$ and form a dense set in $\mathcal{F}_-$, we find that for any $\Psi(0)\in \mathcal{F}_-$, 
the $n$-particle sector of the time-evolved wave function can be obtained by solving the matrix evolution equation
\[
 \partial_t \Psi_n(t) = -\ci H^{(n)}_\lambda \Psi_n(t)\,,
\]
with initial data $\Psi_n(0)$.

\subsection{Dynamics in terms of creation and annihilation operators}\label{sec:CAR}

Antisymmetry of wave vectors is one of the most important features of fermionic quantum systems, and it can alter the properties 
of time-evolution significantly.  Controlling the effect of antisymmetry is difficult in the above formulation of the time-evolution.
A better alternative is offered by representing the time-evolution as an evolution equation of the corresponding fermionic creation and annihilation operators.  We summarize their main properties below and refer to \cite[Section 5.2]{bratteli:ope2} for more mathematical details.

In the present finite lattice case, the Fock space has been constructed using a one-particle space $\mathfrak{h} := \C^{\Lambda}$
and the corresponding (distinguishable) $n$-particle sectors $\mathcal{H}_n :=  \mathfrak{h}^{\otimes n} = \C^{\Lambda^n}$.
Let $P_{-}^{(n)}$ denote the orthogonal projection onto the subspace of antisymmetric functions in $\mathcal{H}_n$; explicitly, 
\[
 (P_{-}^{(n)} \psi)(x_1,\ldots,x_n) =  \frac{1}{n!} \sum_{\pi\in S_n} (-1)^\pi \psi(x_{\pi(1)},\ldots,x_{\pi(n)})\,,
\]
where $S_n$ denotes the group of permutations of the set $\set{1,2,\ldots,n}$ and $(-1)^\pi$ is the sign of the permutation $\pi\in S_n$.
Since we consider a system of identical fermions, at any time, a wave vector $\Psi\in \mathcal{F}_-$
satisfies $P_{-}^{(n)}\Psi_n = \Psi_n$ for all $n$.

Given a one-particle wave vector $g\in \mathfrak{h}$, we define the corresponding \defem{annihilation operator} $a(g)$
as the map which takes a vector 
$\Psi\in \mathcal{F}_-$ and removes the first particle from each of its sectors, with a weight proportional to the overlap with $g$.
More precisely, for a fixed particle number $n\ge 1$, there is a unique bounded linear map $A_n(g):\mathcal{H}_n \to \mathcal{H}_{n-1}$ such that 
for any collection of one-particle wave vectors $f_j\in \mathfrak{h}$,
\[
 A_n(g)\left(\bigotimes_{j=1}^n f_j \right) = \sqrt{n} \braket{g}{f_1} \bigotimes_{j=2}^n f_j \,, 
\]
where $\braket{g}{f}$ is the one-particle scalar product, defined here conjugate linear in the \defem{first} argument, i.e.,
$\braket{g}{f}= \sum_{x\in \Lambda} g(x)^* f(x)$.  
We then define the fermionic annihilation operator $a(g):\mathcal{F}_- \to \mathcal{F}_-$ by the rule
\[
 (a(g)\Psi)_n = P_{-}^{(n)} A_{n+1}(g)  P_{-}^{(n+1)} \Psi_{n+1} = P_{-}^{(n)} A_{n+1}(g) \Psi_{n+1}\,, \qquad n\ge 0\,,\ \Psi \in  \mathcal{F}_- \,.
\]

In general, annihilation operators are unbounded on the appropriate Fock space, and one has to worry about the domain of the operator in its definition. However, it is a remarkable consequence of the antisymmetrisation
that $a(g)$ is in fact a bounded operator on $\mathcal{F}_-$, and the normalisation $\sqrt{n}$ added above guarantees that its operator norm is the same as the norm of the wave vector $g$, i.e., we always have $\norm{a(g)}=\norm{g}_{\mathfrak{h}}$.

The adjoint of $a(g)$, which we denote here by $a^*(g)$, is called the \defem{creation operator} at the vector $g\in \mathfrak{h}$.
The creation operator can indeed be interpreted as creating a particle with wave vector $g$ at the first position (and hence shifting the labels of the existing particles by one).  This
interpretation is based on a more direct construction analogous to the one for $a(g)$ above.  Namely, there is a unique bounded linear map $C_n(g):\mathcal{H}_n \to \mathcal{H}_{n+1}$ such that 
for any collection of one-particle wave vectors $f_j\in \mathfrak{h}$,
\[
 C_n(g)\left(\bigotimes_{j=1}^n f_j \right) = \sqrt{n+1}\, g\otimes f_1 \otimes \cdots \otimes f_n\,, 
\]
We also set $C_0(g)1 = g \in \mathcal{H}_1$.  The fermionic creation operator is then given by 
$c(g):\mathcal{F}_- \to \mathcal{F}_-$, and it satisfies $(c(g)\Psi)_0 = 0$, and 
\[
 (c(g)\Psi)_n = P_{-}^{(n)} C_{n-1}(g)  P_{-}^{(n-1)} \Psi_{n-1} = P_{-}^{(n)} C_{n-1}(g) \Psi_{n-1}\,, \qquad n\ge 1\,,\ \Psi \in  \mathcal{F}_- \,.
\]
One can check that then indeed $c(g)=a^*(g)$ which implies that also $\norm{c(g)}=\norm{g}_{\mathfrak{h}}$.

One important reason why working with the creation and annihilation operators simplifies the analysis of time-evolution is that 
they satisfy fairly simple algebraic rules for swapping the order of any two such operators.  Namely, they satisfy the 
following \defem{canonical anticommutation relations}: for any one-particle vectors $f,g\in \mathfrak{h}$, we have 
\begin{eqnarray}\label{eq:canonicalac}
 & a(f) a(g) + a(g)a(f) = 0 = a(f)^* a(g)^* + a(g)^* a(f)^*\,, \nonumber \\
 & a(f) a(g)^* + a(g)^* a(f) = \braket{f}{g}1\,,  
\end{eqnarray}
where ``$1$'' denotes the identity operator on $\mathcal{F}_-$.  In particular, $a(f)^2 = 0 = a^*(f)^2$, and if $(e_\ell)$ is any orthonormal basis of $\mathfrak{h}$, we have
\[
 a(e_\ell) a(e_{\ell'})^* + a(e_{\ell'})^* a(e_{\ell}) = \cf{\ell = \ell'} 1\,,
\]
with $\cf{P}$ denoting the \defem{generic characteristic function} of the condition $P$: we define $\cf{P}=1$, if $P$ is true,
and $\cf{P}=0$, if $P$ is false.

Moreover, tensor products in $\mathcal{H}_n$ are conveniently expressed in terms of products of creation operators acting on the vacuum 
$\Omega=(1,0,0,\ldots)\in \mathcal{F}_-$.  Namely, if $g_j\in \mathfrak{h}$, $j=1,2,\ldots,n$, are given,
then $\otimes_j g_j \in \mathcal{H}_n$ after antisymmetrisation defines a vector $\Psi \in \mathcal{F}_-$ by setting all other components to zero,
i.e., setting
$\Psi_n = P_{-}^{(n)}( \otimes_j g_j)$ and $\Psi_m=0$, for $m\ne n$.  This vector can also be obtained from
\begin{align}\label{eq:projecttensorpv}
 \Psi = \frac{1}{\sqrt{n!}} a^*(g_1)\cdots a^*(g_n)\Omega\,. 
\end{align}

The collection of creation and annihilation operators corresponding to the standard unit vector orthonormal basis $(e_x)_{x\in \Lambda}$, where $(e_x)_y=\cf{x=y}$ for all $x,y\in \Lambda$, is of particular interest to us. 
We employ the following standard shorthand notations:
\begin{align}\label{eq:defax}
 a(x) := a(e_x)\,,\quad a^*(x) := a^*(e_x) = a(x)^*\,, \qquad x\in \Lambda\,.
\end{align}
These operators can be thought of as annihilating or creating a particle at the site $x$.
By (\ref{eq:canonicalac}), they satisfy the following simple anticommutation relations for any $x,y\in \Lambda$,
\begin{eqnarray}\label{eq:axantic} 
 & a(x) a(y) + a(x)a(y) = 0 = a(x)^* a(y)^* + a(x)^* a(y)^*\,,\nonumber \\
 & a(x) a(y)^* + a(x)^* a(y) = \cf{x=y}1\,.
\end{eqnarray}

We can also use the creation operators to generate an \defem{orthonormal basis} for $\mathcal{F}_-$.
For this, first define
\[
 e(x_1,\ldots,x_n) := a^*(x_1)\cdots a^*(x_n)\Omega\,, \qquad x_i\in \Lambda\,,\ i=1,2,\ldots,n\,.
\]
The orthonormal basis may be constructed by collecting all non-repeating sequences of arbitrary length
and then choosing one representative for each collection of sequences which differ by a permutation of particle labels.
The actual choice does not does not play much role: if $(x_i)\in \Lambda^n$ and $\pi \in S_n$ is some permutation, then by the anticommutation relations
\[
  e(x_{\pi(1)},\ldots,x_{\pi(n)}) = (-1)^\pi  e(x_1,\ldots,x_n)\,,
\]
and hence the choice merely affects signs of the basis vectors.

After these preliminaries, it is straightforward to check that wave vectors and interaction potentials may also be represented using the creation and annihilation operators.  Namely, if $\Psi\in \mathcal{F}_-$, $n\in \N$, and $x\in \Lambda^n$, we have
\[
 \Psi_n(x_1,\ldots,x_n) = \braket{\otimes_{i=1}^n e_{x_i}}{\Psi_n}_{\mathcal{H}_n}
 = \braket{P_{-}^{(n)}(\otimes_{i=1}^n e_{x_i})}{\Psi_n}_{\mathcal{H}_n}\,,
\]
and hence by (\ref{eq:projecttensorpv}),
\[
 \Psi_n(x_1,\ldots,x_n) = \frac{1}{\sqrt{n!}} \braket{a^*(x_1)\cdots a^*(x_n)\Omega}{\Psi}_{\mathcal{F}_-}\,.
\]
Moreover, the anticommutation relations imply that if $x,y$ and $x_i\in \Lambda$, $i=1,2,\ldots,n$, then 
\begin{align}\label{eq:basicasa}
 & a^*(x) a(y) a^*(x_1)\cdots a^*(x_n)\Omega \nonumber\\
 & \quad = \sum_{i=1}^n \cf{y=x_i}a^*(x_1)\cdots a^*(x_{i-1}) a^*(x)
 a^*(x_{i+1}) \cdots a^*(x_n)\Omega\,. 
\end{align}
Using these two properties it is now straightforward to check that the earlier defined operators $H_0$ and 
$V$ on the fermionic Fock space have the following representations in terms of creation and annihilation operators,
\begin{align}
 H_0 & = \sum_{x,y\in \Lambda} \alpha(x-y;L) a(x)^* a(y)\,,\\
 V & = \frac{1}{2}\sum_{x,y\in \Lambda} V(x-y;L) a(x)^* a(y)^* a(y) a(x)\,. 
\end{align}
The above right hand sides are finite sums in the Banach space of bounded operators on $\mathcal{F}_-$, and thus $H_0$, $V$,
and $H_\lambda= H_0 + \lambda V$ are also bounded operators on the fermionic Fock space.

The time-evolution of any initial data $\Psi(0)\in \mathcal{F}_-$ under the 
semigroup $U_t := \rme^{-\ci t H_\lambda} $ can be solved if we can solve the time-evolution of the annihilation operators, i.e., it suffices to study 
\[
 a(x,t) := \rme^{\ci t H_\lambda} a(x) \rme^{-\ci t H_\lambda} \,,
\]
and its adjoint
\[
 a^*(x,t) := \rme^{\ci t H_\lambda} a^*(x) \rme^{-\ci t H_\lambda} \,.
\]
This follows from our definition that the Hamiltonian acts trivially on the vacuum sector, $(H_\lambda)_0=0$,
and thus
\[
 a^*(x_1,t)\cdots a^*(x_n,t)\Omega = \rme^{\ci t H_\lambda} a^*(x_1)\cdots a^*(x_n)\Omega\,,
\]
implying that
\begin{align*}
& \Psi_n(x_1,\ldots,x_n,t) = \frac{1}{\sqrt{n!}} \braket{a^*(x_1)\cdots a^*(x_n)\Omega}{\rme^{-\ci t H_\lambda} \Psi(0)}_{\mathcal{F}_-} \\
& \quad =\frac{1}{\sqrt{n!}} \braket{a^*(x_1,t)\cdots a^*(x_n,t)\Omega}{\Psi(0)}_{\mathcal{F}_-}\,. 
\end{align*}

Since the Hamiltonian is a bounded operator, we can directly differentiate the definition and obtain 
\[
 \partial_t a(x,t) = -\ci \rme^{\ci t H_\lambda} [a(x),H_\lambda] \rme^{-\ci t H_\lambda} \,.
\]
The computation of the commutator is straightforward using the anticommutation relations, yielding
\[
  [a(x),H_\lambda] = \sum_{y\in \Lambda} \alpha(x-y;L) a(y) + \lambda \sum_{y\in \Lambda} V(x-y;L)  a(y)^* a(y) a(x)\,.
\]
Therefore, we find that in order to solve the original (linear) evolution equation in the fermionic Fock space,
it suffices to solve the following non-linear operator evolution equation on the space of bounded operators on $\mathcal{F}_-$,
\begin{align}
  \partial_t a(x,t) = -\ci \sum_{y\in \Lambda} \alpha(x-y;L) a(y,t)
  -\ci \lambda \sum_{y\in \Lambda} V(x-y;L)  a^*(y,t) a(y,t) a(x,t)\,.
\end{align}
In Fourier variables, after defining
\[
 \FT{a}(k,t) := \sum_{x\in \Lambda} \rme^{-\ci 2\pi x \cdot k} a(x,t)\,,
\]
we obtain
\begin{align}
  & \partial_t \FT{a}(k,t) 
  = -\ci \omega(k) \FT{a}(k,t) \nonumber  \\ & \quad 
  -\ci \lambda \int_{(\Lambda^*)^3} \! \rmd k_1 \rmd k_2\rmd k_3\,
  \delta_{\Lambda}(k-k_1-k_2-k_3) \FT{V}(k_1+k_2) \FT{a^*}(k_1,t) \FT{a}(k_2,t) \FT{a}(k_3,t) \,,
\end{align}
where $\delta_\Lambda(k) := |\Lambda| \cf{k=0 \bmod \Lambda^*}$ is a ``discrete Dirac $\delta$-function''
and $[\FT{a}(k,t)]^*= \FT{a^*}(-k,t)$.

\subsection{Fermionic systems with spin interactions and the Hubbard model}\label{sec:spin}

Spin is an integral part of description of quantum mechanical particles.  For instance,
by the spin-statistics relation, all fermionic particles possess a half-integer spin. 
In particular, the spin cannot be zero, so the above fermionic description is not yet completely adequate for physical fermions.

Spin is a one-particle property, and hence affects the definition of 
the one-particle Hilbert space $\mathfrak{h}$ above.  It is determined by a half-integer value 
$S\in \N_0/2$, resulting in $2 S+1$ new ``internal'' degrees of freedom which are labelled
by values in $\sigma_S := \set{-S,-S+1,\ldots,S}$.  There are several equivalent ways of 
defining the wave vector of a particle with a non-zero spin: one can either think that they are multicomponent
wave-vectors, $\psi(x)\in \C^{\sigma_S}$, or that each lattice site is augmented with $D$ extra degrees of freedom,
$\psi(x,\sigma)\in \C$, $\sigma \in \sigma_S$.  These descriptions are quantum mechanically equivalent  since the identification 
\[
 \psi(x)_\sigma =  \phi(x,\sigma) = \braket{e_x\otimes e_\sigma}{\phi}
\]
provides a mapping $\psi\to \phi$ which turns out to be a Hilbert space isomorphism between $\oplus_{\sigma \in \sigma_S} L^2(\Lambda)$ and $L^2(\Lambda \times \sigma_S)$.  The second equality above also yields an isomorphism, namely the standard one between $L^2(\Lambda \times \sigma_S)$ and $L^2(\Lambda) \otimes L^2(\sigma_S)$.

Hence, most of the discussion in the previous sections holds verbatim if we replace $x\in \Lambda$ by 
$(x,\sigma)\in \Lambda \times \sigma_S$.  The main differences come from the physical restrictions for the spin-interactions which have no need to be ``translation invariant'' in the spin-degrees of freedom.  Thus Fourier-transforming the spin-degrees is not helpful and, instead, one should try to aim at simplifications by
finding other unitary transformations which diagonalise at least part of the Hamiltonian.

One case which reduces to the discussion without spin, occurs when the total Hamiltonian $H$ can be diagonalised with respect to the spin degrees of freedom, i.e., if there is a unitary transformation $U$ for which
$U^* H U = \oplus_{\sigma \in \sigma_S} H_\sigma$.  Then after the unitary transformation each spin-component 
evolves independently from the others and thus it satisfies the ``spinless'' equations of the previous section.

Spatially translation invariant generalisations of the previous weakly interacting Hamiltonians are determined by the operators
\begin{align}\label{eq:Hwithspin}
 H_0 & = \sum_{x,y\in \Lambda} \sum_{\sigma,\sigma'\in \sigma_S}  \alpha_{\sigma\sigma'}(x-y;L) a(x,\sigma)^* a(y,\sigma')\,,\\
 \label{eq:Vwithspin}
 V & = \frac{1}{2}\sum_{x,y\in \Lambda} \sum_{\sigma,\sigma'\in \sigma_S} V_{\sigma\sigma'}(x-y;L) a(x,\sigma)^* a(y,\sigma')^* a(y,\sigma') a(x,\sigma)\,.
\end{align}
The functions $\alpha_{\sigma\sigma'}(x;L)$ and $V_{\sigma\sigma'}(x;L)$ are constructed as in (\ref{eq:defalphaL}) and (\ref{eq:defVL}),
using some given $\omega_{\sigma\sigma'}:\T^d\to \R$ and $\FT{V}_{\sigma\sigma'}:\T^d\to \R$,
for each $\sigma,\sigma'$.
We require $H_0$ to be self-adjoint and the interaction symmetric under spatial inversions, and this is guaranteed by 
assuming that $\omega(-k)=\omega(k)=\omega(k)^*$, as $S\times S$ -matrices.
Similarly, the self-adjointness of $V$ can be guaranteed by assuming that each $\FT{V}(k)$ is a Hermitian matrix
and that they satisfy an additional symmetry property $\FT{V}_{\sigma\sigma'}(-k) = \FT{V}_{\sigma'\sigma}(k)$ related to particle permutation invariance.

One well studied example of this type is the \defem{Hubbard model} which concerns spin-$\frac{1}{2}$ fermions like electrons.  Then $S=\frac{1}{2}$ and usually one simplifies the discussion by labelling the spin degrees of freedom $\set{-\frac{1}{2},\frac{1}{2}}$ using the sign, i.e., using the set $2\sigma_S = \set{-1,1}$ for labelling.
In the Hubbard model the free evolution 
is taken to be fully spin rotation invariant,
\begin{align}
 H_0 & = \sum_{x,y\in \Lambda}\sum_{\sigma=\pm 1}  \alpha(x-y;L) a(x,\sigma)^* a(y,\sigma)\,,
\end{align}
and thus
depending only on one dispersion relation function which is typically chosen to be nearest neighbour,
$\omega(k) = -\sum_{\nu =1}^d \cos(2\pi k_\nu)$.  The pair interactions in the Hubbard model are taken to be onsite only,
\begin{align}
 V & = \frac{1}{2}\sum_{x\in \Lambda} \sum_{\sigma,\sigma'=\pm 1} V_{\sigma\sigma'} a(x,\sigma)^* a(x,\sigma')^* a(x,\sigma') a(x,\sigma)\,,
\end{align}
and, since $a(x,\sigma)^2=0$ and $V_{\sigma\sigma'}=V_{\sigma\sigma'}(0;L)$, $\sigma,\sigma'\in \set{\pm 1}$, form a real symmetric $2\times 2$ matrix, without loss of generality, we may set $V_{\sigma\sigma}=0$ and use $V_{+-}=V_{-+}$ as the sole real parameter.  It is usually included in the definition of the coupling $\lambda$, and thus the general fermionic spin-$\frac{1}{2}$ onsite interactions are covered by the interaction\footnote{The most standard notation for the Hubbard model uses the 
potential $U\sum_x n(x,+) n(x,-)$ where $n(x,\sigma):=a(x,\sigma)^* a(x,\sigma)$.  This is seen to be equivalent to the present case after setting $U=\lambda$ and using the anticommutation relations.}
\begin{align}
 V_{\text{Hubbard}} & = \sum_{x\in \Lambda} a(x,+)^* a(x,-)^* a(x,-) a(x,+)\,.
\end{align}

Let us point out that onsite potentials fall into the class of translation invariant potentials studied in the previous subsection. Namely, they correspond to choosing potentials whose Fourier transforms are constant, $\FT{V}_{\sigma\sigma'}(k)=V_{\sigma\sigma'}$ for all $k\in \T^d$.

The main difficulties compared to deriving the evolution equations in the earlier discussed case are notational.
We skip the parts which are similar to the earlier computations, and merely record the outcome in a form which 
is easy to use in computations involving products of creation and annihilation operators.

We label annihilation operators with an additional label $\tau=-1$ and creation operators with $\tau=+1$,
and consider their dynamics after Fourier transform of the spatial degrees of freedom.
Explicitly, we define
\begin{align}
 a(k,\sigma,-1,t) & := \FT{a}(k,\sigma,t) = \sum_{x\in \Lambda} \rme^{-\ci 2\pi x \cdot k} a(x,\sigma,t)\,,\\
 a(k,\sigma,+1,t) & := \FT{a^*}(k,\sigma,t) = \sum_{x\in \Lambda} \rme^{-\ci 2\pi x \cdot k} a^*(x,\sigma,t)\,.
\end{align}
These operators are connected via operator adjoints,
$[a(k,\sigma,\tau,t)]^* = a(-k,\sigma,-\tau,t)$.
Since now
\begin{align*}
 &   \partial_t a(x,\sigma,t) = -\ci \sum_{x'\in \Lambda}\sum_{\sigma'\in \sigma_S}  \alpha_{\sigma\sigma'}(x-x';L) a(x',\sigma',t)
 \\&\quad -\ci \lambda \sum_{x'\in \Lambda}\sum_{\sigma'\in \sigma_S} V_{\sigma\sigma'}(x-x';L)  a^*(x',\sigma',t) a(x',\sigma',t) a(x,\sigma,t)\,,
\end{align*}
the above operators satisfy the following closed evolution equations
\begin{align}\label{eq:atautimeder}
  & \partial_t a(k,\sigma,\tau,t) 
  = \ci \tau \sum_{\sigma'\in \sigma_S}  \omega_{\sigma\sigma'}(k;\tau) a(k,\sigma',\tau,t) \nonumber 
   \\ &  \qquad
  + \ci \tau \lambda \sum_{\sigma_1,\sigma_2,\sigma_3\in \sigma_S} \int_{(\Lambda^*)^3} \! \rmd k_1 \rmd k_2\rmd k_3\,
  \delta_{\Lambda}(k- k_1-k_2 - k_3) 
  \nonumber  \\ & \qquad\quad  \times \FT{V}_{\sigma,\sigma_1,\sigma_2,\sigma_3}(k_1,k_2,k_3;\tau) a(k_1,\sigma_1,1,t) a(k_2,\sigma_2,\tau,t) 
  a(k_3,\sigma_3,-1,t) \,,
\end{align}
where $\omega_{\sigma\sigma'}(k;-1):=\omega_{\sigma\sigma'}(k)$, $\omega_{\sigma\sigma'}(k;+1):=\omega_{\sigma'\sigma}(k)$, and
\begin{align*}
 & \FT{V}_{\sigma,\sigma_1,\sigma_2,\sigma_3}(k_1,k_2,k_3;-1) = 
 \cf{\sigma_1=\sigma_2,\sigma_3=\sigma} \FT{V}_{\sigma\sigma_2}(k_1+k_2) \,,\\
 & \FT{V}_{\sigma,\sigma_1,\sigma_2,\sigma_3}(k_1,k_2,k_3;+1) = 
 \cf{\sigma_1=\sigma,\sigma_3=\sigma_2} \FT{V}_{\sigma\sigma_2}(k_2+k_3) \,.
\end{align*}

Here we need the above equations only in two special cases.
First, if there is no spin, the equation reduces to 
\begin{align}\label{eq:atauspinless}
  & \partial_t a(k,\tau,t) 
  = \ci \tau  \omega(k) a(k,\tau,t) \nonumber 
  + \ci \tau \lambda \int_{(\Lambda^*)^3} \! \rmd k_1 \rmd k_2\rmd k_3\,
  \delta_{\Lambda}(k- k_1-k_2 - k_3) 
  \nonumber  \\ & \qquad\quad  \times 
  \FT{V}(k_1,k_2,k_3;\tau) a(k_1,1,t) a(k_2,\tau,t) 
  a(k_3,-1,t) \,,
\end{align}
with $\FT{V}(k_1,k_2,k_3;-1)=\FT{V}(k_1+k_2)$ and $\FT{V}(k_1,k_2,k_3;1)=\FT{V}(k_2+k_3)$.
Secondly, for the Hubbard model, the equations can be simplified into 
\begin{align}\label{eq:atauHubbard}
  & \partial_t a(k,\sigma,\tau,t) 
  = \ci \tau \omega(k) a(k,\sigma,\tau,t) \nonumber 
  + \ci \tau \lambda \int_{(\Lambda^*)^3} \! \rmd k_1 \rmd k_2\rmd k_3\,
  \delta_{\Lambda}(k- k_1-k_2 - k_3) 
  \nonumber  \\ & \qquad\quad  \times 
  a(k_1,\tau \sigma,1,t) a(k_2,-\sigma,\tau,t) a(k_3,-\tau \sigma,-1,t) \,.
\end{align}

\section{States, reduced density matrices, and truncated correlation functions}\label{sec:states}

A state in classical mechanics is a probability measure describing the distribution of positions and velocities
of the particles at some fixed time.  Thus it can be used to compute the statistics of all observables,
i.e., measurable functions of the positions and velocities at that time.   
In Hamiltonian mechanics, an initial state given at time $t=0$ determines the state at all times $t\in \R$.
Often it is simpler to study the evolution of physical properties of the system by inspecting the evolution 
starting from some suitably chosen random initial state rather than from a deterministic state with fixed values for the initial positions and velocities of the particles.

A \defem{state} at time $t$ in quantum mechanics is defined as a map $\rho_t$ which associates to each observable $A$
a number $\rho_t[A]$ which gives the limiting value for statistical averages of this observable measured in repeated
experiments.  This is analogous to the expectation value map under the probability measure which defines the state in the classical case.  The more precise mathematical definition of a state takes two ingredients:
the collection of observables $\mathcal{A}$, which is assumed to be some subspace of bounded operators, closed under adjoint and containing the identity operator, and a positive linear functional $\rho:\mathcal{A}\to \C$ of norm $1$.

For instance, a Borel probability measure $\mu$ of  wave vectors $\Psi\in \mathcal{H}$, $\norm{\Psi}=1$, generates a state
by setting for any bounded operator $A$ on $\mathcal{H}$
\[
 \rho[A] := \int\! \mu(\rmd \psi) \, \braket{\psi}{A\psi}\,.
\]
Most often a state is determined by giving a trace-class operator $\rho$ on $\mathcal{H}$ such that 
$\rho$ is positive, $\tr \rho=1$, and setting $\rho[A]=\tr[\rho A]$ for all $A\in \mathcal{A}$.  Such an operator $\rho$ is called the \defem{density matrix} of the state (note that we do not make a distinction in the notation between the state and its density matrix).  If the Hilbert space is separable, such as our Fock spaces are, then for instance all states given by the above Borel probability measures have a density matrix associated with them.

The $n$:th \defem{reduced density matrix} $\rho_n$ 
is an analogous quantity which is obtained from the full density matrix
by taking a partial trace over the degrees of freedom which concern particle labels higher than $n$.
The general construction is discussed in \cite[Section 6.3.3]{bratteli:ope2} and in \cite[Section 3]{ls09}, but
there is a more direct definition available for the present system of lattice fermions: Given a state $\rho$ on 
the fermionic Fock space, we first define
\begin{align}\label{eq:defreddensmatr}
 \rho_n(z_1,z'_1,\ldots,z_n,z'_n) := \rho[a^*(z'_1)\cdots a^*(z'_n)a(z_n)\cdots a(z_1)]\,.
\end{align}
Here each $z_i$ and $z'_i$ belongs to the one-particle label set, i.e., $z_i\in \Lambda$ in the spinless case
and $z_i\in \Lambda\times \sigma_S$ for spin-$S$ particles.  The collection of these complex numbers
defines the reduced density matrix $\rho_n$, which is a positive operator on $\mathfrak{h}^{\otimes n}$,
via the formula
\[
 \braket{\otimes_i z_i}{\rho_n (\otimes_i z'_i)} = \rho_n(z_1,z'_1,\ldots,z_n,z'_n)\,.
\]

In quantum mechanics, given an initial density matrix $\rho(0)=\rho$, the expectation of a
time-evolved observable $A(t) = U_t^* A U_t$ satisfies
\[
 \rho[A(t)] = \tr[\rho U_t^* A U_t] = \tr[ U_t\rho U_t^* A] \,,
\]
by cyclicity of trace.  Hence, we define the time-evolved density matrix $\rho(t):=U_t\rho U_t^*$
for which $\rho(t)[A]=\rho[A(t)]$.  The reduced time-evolved  density matrices may thus be obtained as
expectations of time-evolved creation and annihilation operators: by replacing each $a(z)$ 
in (\ref{eq:defreddensmatr}) by $a(z,t)=U_t^* a(z) U_t$, we obtain the reduced density matrix $\rho(t)_n$.

Considering the earlier observation that time-evolved annihilation operators suffice to determine the time-evolution of
wave vectors, it is not surprising that reduced density matrices play an important role in the physics of quantum fluids.  For instance, the expectation of the hopping Hamiltonian $H_0$ may be computed from $\rho(t)_1$ by the 
formula
\[
 \rho(t)[H_0] = \sum_{x,y\in \Lambda} \sum_{\sigma,\sigma'\in \sigma_S}  \alpha_{\sigma\sigma'}(x-y;L) 
 \rho(t)_1((x,\sigma),(y,\sigma'))\,.
\]
Indeed, for kinetic theory, the central goal is to describe the evolution of $\rho(t)_1$, a positive operator on $\mathfrak{h}$, in the limit of weak coupling.

In fact, there is a class of fermionic states, called \defem{quasifree states}, for which $\rho_1$ uniquely determines all other reduced density matrices: if $\rho$ is quasifree, then for all $n\ge 1$ the corresponding 
density matrix is given as a determinant of an $n\times n$ matrix,
\[
 \rho_n(z_1,z'_1,\ldots,z_n,z'_n) = \det (\rho_1(z_i,z'_j))_{i,j=1,\ldots ,n}\,.
\]
To simplify analysis of states which are not quasifree but close to such, one can introduce 
\defem{truncated correlation functions} $\rho^T$ which are analogous to cumulants of random variables in classical probability theory.  The construction below applies to a state $\rho$ on a fermionic system which is 
\defem{even}: it is assumed that an expectation of any observable remains invariant if we change $a(z)$ to $-a(z)$ for all $z$.
As explained in more detail in \cite[pp.\ 42--43]{bratteli:ope2}, given 
an even state $\rho$ to each even length sequence $(a_1,a_2,\ldots,a_m)$ 
of creation and annihilation operators one may 
associate a truncated expectation $\rho^T[a_1,a_2,\ldots,a_m]$ such that the expectation of
any product of even length can be expressed as a sum over partitions.  Explicitly,
\begin{align}\label{eq:momtotrunc}
 \rho[a^I] = \sum_{\Pi\in \mathcal{P}_2(I)} \vep(\Pi) \prod_{S\in \Pi} \rho^T[a_S]\,,
\end{align}
where $I=(1,2,\ldots,n)$, $a^I:=a_1 \cdots a_n$,
$\mathcal{P}_2(I)$ denotes the collection of partitions of $I$ into even length subsequences,
$\vep(\Pi)$ is the sign of the permutation which takes $I$ to $\Pi=(S_1,\ldots,S_m)$, and for a subsequence
$S=(s_1,\ldots,s_m)$ of $I$ we have used the shorthand notation $a_S=(a_{s_1},\ldots,a_{s_m})$.
Note that odd sequences for even states have always zero expectation, so this is the antisymmetrised analogue of 
the moments-to-cumulants formula of probability. 

The above definition requires careful consideration of the signs of each term.
The following identity can also serve as a basis for a recursive definition of the truncated expectations,
\begin{align}\label{eq:iterdeftrunc}
 \rho[a^I] = \sum_{m\in S\subset I} \vep(S,I\setminus S) \rho^T[a_S] \rho[a^{I\setminus S}]\,, 
\end{align}
where $m\in I$ is any fixed label and
$\vep(S,I\setminus S)$ is the sign of the permutation $I\to (S,I\setminus S)$.  (Note that all terms where $S$ has an odd length are zero in the sum, since then also $I\setminus S$ is odd, so we could have restricted the sum to even subsequences here.)
For instance, $\rho^T[a_1,a_2] = \rho[a_1 a_2]$, and 
for $n=4$ we have
\begin{align*}
  & \rho[a_1 a_2 a_3 a_4]  = \rho^T[a_1,a_2,a_3,a_4] 
  \\ & \quad
 + \rho^T[a_1,a_2] \rho[a_3 a_4]
 -  \rho^T[a_1,a_3] \rho[a_2 a_4] +\rho^T[a_1,a_4] \rho[a_2 a_3]\,, 
\end{align*}
and thus
\begin{align*}
  & \rho^T[a_1,a_2,a_3,a_4]   := \rho[a_1 a_2 a_3 a_4]
  \\ & \quad
  - \rho[a_1 a_2] \rho[a_3 a_4]
  +  \rho[a_1 a_3] \rho[a_2 a_4] -\rho[a_1 a_4] \rho[a_2 a_3]\,, 
\end{align*}
and, in accordance with (\ref{eq:momtotrunc}), also 
\begin{align}\label{eq:tr4tomom}
  & \rho[a_1 a_2 a_3 a_4]  = \rho^T[a_1,a_2,a_3,a_4] 
 \nonumber \\ & \quad
 + \rho^T[a_1,a_2] \rho^T[a_3, a_4]
 -  \rho^T[a_1,a_3] \rho^T[a_2, a_4] +\rho^T[a_1,a_4] \rho^T[a_2, a_3]\,.
\end{align}

The truncated correlation functions can be used to characterise quasifree states:
\defem{an even state $\rho$ is quasifree if and only if $\rho^T[a_1,a_2,\ldots,a_n]=0$ for all $n>2$.\/}
This is completely analogous with characterisation of Gaussian measures by vanishing of their higher order cumulants.
Even for states which are not quasifree,  the truncated correlation functions enjoy properties which are typically not valid for direct expectations:
\begin{enumerate}
 \item If $n>2$, then $\rho^T[a_1,a_2,\ldots,a_n]$ is completely antisymmetric with respect to permutation of its arguments: if $\pi\in S_n$, we have $\rho^T[a_{\pi(1)},a_{\pi(2)},\ldots,a_{\pi(n)}]=(-1)^\pi \rho^T[a_1,a_2,\ldots,a_n]$. 
 (For a proof, consider a basic odd permutation which swaps two neighbouring labels $m$ and $m'$, 
 and then then use (\ref{eq:iterdeftrunc}) and the anticommutation relations.)
 \item\label{it:ell1corr} If $\rho$ is an equilibrium Gibbs state at sufficiently small activity and corresponding to a short range interaction, 
 all reduced density matrices are typically decaying summably in the separation of their spatial arguments. 
 For a precise statement and assumptions under which this result holds, see 
 \cite[Theorem 6.3.21]{bratteli:ope2}, and further discussion can be found in \cite{Salm08}.  In particular, keeping one of the sites fixed, Fourier transforms of the reduced density matrices are typically uniformly bounded in the lattice size $L$, unlike those of the corresponding expectations.
\end{enumerate}

\section{Weak coupling limit and quantum kinetic theory}

For kinetic theory, we are interested in the evolution of the first truncated reduced density matrix $\rho_1(x',\sigma',x,\sigma;t)=\rho^T[a^*(x',\sigma',t), a(x,\sigma,t)]$.  There is no difference between the truncated and direct reduced density matrices for the first reduced density matrix of an even state of fermions but for higher order density matrices there is a difference in their properties.  Most notably, for systems which which are eventually well approximated by Gibbs states of the type discussed in item \ref{it:ell1corr} at the end of Section \ref{sec:states}, one would expect the \defem{truncated} correlation functions to decay in the distance.  Then, Fourier transforms in these variables are given by ``nice'' functions, for instance, uniformly bounded in the lattice size or with a uniformly bounded $L^2(\rmd k)$-norm.  In contrast, the Fourier transform of the corresponding moments would be a fairly complicated sum over ``$\delta_\Lambda$-distributions''.

Here we consider only initial data which are both \defem{gauge invariant} and \defem{translation invariant}.
The first condition means that the initial data does not contain correlations between different particle sectors, and 
this property is preserved by the present type of evolution.  It simplifies the resulting analysis since
for gauge invariant states 
all moments, which do not have the same number of creation and annihilation operators,
are zero.  For instance, then $ \rho[a(y,\sigma'',t)a(x,\sigma,t)]=0=\rho[a^*(y,\sigma'',t)a^*(x,\sigma,t)]$.

For translation invariance, we require that all moments are invariant under periodic spatial translations
of the lattice $\Lambda$.  For the present translation invariant $H_0$ and $V$ also this property is preserved by 
the time-evolution.  
As a consequence, any one of the spatial arguments of the correlation functions can be translated to the origin.  
In particular, 
there is a function $F:\Lambda\times\R \to \C^{2\times 2}$ for which 
\[
 \rho_1(x',\sigma',x,\sigma;t) = F_{\sigma'\sigma}(x'-x,t)\,.
\]
The \defem{Wigner function} is defined as 
the discrete Fourier transform of $F$,
\begin{align}\label{eq:defWigner}
 & W_{\sigma'\sigma}(k,t) := \sum_{y\in\Lambda} \rme^{-\ci 2\pi y\cdot k}F_{\sigma'\sigma}(y,t)
 =
  \int_{\Lambda^*} \!\rmd k'\, \rho[a(k,\sigma',1,t) a(k',\sigma,-1,t)]
  \,.
\end{align}
Using the properties of adjoints, it is straightforward to check that the so defined $\sigma_S\times \sigma_S$
matrix $W(k,t)$ is always Hermitian.  In addition, translation invariance may be invoked to prove that
\begin{align}\label{eq:aatoWigner}
\rho[a(k,\sigma',1,t) a(k',\sigma,-1,t)] = W_{\sigma'\sigma}(k,t) \delta_\Lambda(k+k')\,.
\end{align}

We also introduce the related notation $\tilde{W}$ for the corresponding expectation where the order of 
the operators has been swapped.  More precisely,
we define as matrices
\begin{align}\label{eq:deftildeWigner}
 \tilde{W}(k,t) := 1-W(k,t)\,,
\end{align}
where $1$ denotes the diagonal unit matrix.
By the anticommutation relations, then 
\begin{align}\label{eq:aatotildeWigner}
\rho[ a(k',\sigma,-1,t)a(k,\sigma',1,t)] = \tilde{W}_{\sigma'\sigma}(k,t) \delta_\Lambda(k+k')\,.
\end{align}

The quantum kinetic equation will concern the time-evolution of the above Hermitian 
matrix-valued Wigner functions.  
There are a number of differences in the computations depending on whether there
are spin-interactions present or not, and we have split the discussion accordingly below.

\subsection{Fermionic Boltzmann--Nordheim equation}

We begin with a case in which the spin-degrees of freedom evolve independently.
As mentioned above, this case can be handled ignoring the spin degrees of freedom and 
thus we can use the spinless results and notations.
We adapt here the method introduced in \cite{LM16} for derivation of a phonon Boltzmann
equation for the weakly nonlinear discrete Schr\"odinger equation from the evolution hierarchy 
of \defem{truncated} correlation functions.  For comparison,
a derivation of the Boltzmann--Nordheim equation
using direct perturbation expansions of moments
and their graph representations can be found in \cite{ls09}.  

It should be stressed 
that neither method currently produces a mathematically rigorous derivation of fermionic 
kinetic theory.  In particular, it is not yet known which precise assumptions are needed for the kinetic approximation
to work nor are there any rigorous bounds for the accuracy of the approximation.
From the point of view of the truncated correlation function hierarchy, the key missing ingredient 
is a control of the evolution of decay properties of correlation functions.  Here we do not go into any detail 
about the role played by the terms ignored in the derivations below but more details about why
their effects are in general expected to be lower order in the weak coupling limit
$\lambda \to 0$ can be found in \cite{LM16,ls09}.   

Let us also point out one case in which rigorous control
has been possible:  in \cite{NLS09}, the kinetic scaling limit of time-correlations of equilibrium distributed fields 
with discrete nonlinear Schr\"odinger evolution are proven to follow the above scenario.  In this case,
the state itself is stationary and the good decay properties of the truncated correlation functions
are provided by the initial data which can be studied with methods from equilibrium statistical mechanics.

Differentiating (\ref{eq:defWigner}) and recalling the adjoint relations yields the following 
representation for the time derivative of the Wigner function of translation invariant states
\begin{align}\label{eq:Wevol}
 & \partial_t  W_{\sigma'\sigma}(k,t) 
 \nonumber \\&\quad
 =  \int_{\Lambda^*} \!\rmd k'\, 
 \left(\rho[\partial_t a(k,\sigma',1,t) a(k',\sigma,-1,t)]+
 \rho[\partial_t a(-k',\sigma,1,t) a(-k,\sigma',-1,t)]^*\right)
 \nonumber \\&\quad
 =  \int_{\Lambda^*} \!\rmd k'\, 
 \left(\rho[\partial_t a(k,\sigma',1,t) a(k',\sigma,-1,t)]+
 \rho[\partial_t a(k,\sigma,1,t) a(k',\sigma',-1,t)]^*\right)\,.
\end{align}
Thus for a translation invariant states of fermions without spin, we have
\begin{align}\label{eq:spinlessWevol}
 & \partial_t  W(k,t) 
 = 2\re\left( \int_{\Lambda^*} \!\rmd k'\, 
 \rho[\partial_t a(k,1,t) a(k',-1,t)]\right)\,.
\end{align}
We use (\ref{eq:atauspinless}) to compute the derivative, yielding
\begin{align}\label{eq:spinlder}
& \int_{\Lambda^*} \!\rmd k'\, 
 \rho[\partial_t a(k,1,t) a(k',-1,t)]
 = \ci \omega(k) \int_{\Lambda^*} \!\rmd k'\, 
 \rho[a(k,1,t) a(k',-1,t)]
   \nonumber \\ & \qquad +
  \ci \lambda \int_{(\Lambda^*)^4} \! \rmd k_1 \rmd k_2\rmd k_3\rmd k_4\,
  \FT{V}(k_2+k_3)\delta_\Lambda( k-k_1-k_2-k_3)
   \nonumber \\ & \qquad\quad  \times 
   \rho[a(k_1,1,t) a(k_2,1,t) a(k_3,-1,t) a(k_4,-1,t)]\,.
\end{align}
The first term on the right is purely imaginary and does not contribute to the real part.
In the second term, the expectation is antisymmetric with respect to the swap $k_1\leftrightarrow k_2$, and thus
we can conclude that
\begin{align}\label{eq:spinlessWevol2}
 & \partial_t  W(k,t) 
 = \re\Bigl[\ci \lambda \int_{(\Lambda^*)^4} \! \rmd k_1 \rmd k_2\rmd k_3\rmd k_4\,
   \left(\FT{V}(k_2+k_3)-\FT{V}(k_1+k_3)\right)
   \nonumber \\ & \qquad  \times 
  \delta_\Lambda( k-k_1-k_2-k_3)
 \rho[a(k_1,1,t) a(k_2,1,t) a(k_3,-1,t) a(k_4,-1,t)]\Bigr]
 \,.
\end{align}

We represent the remaining expectation in terms of truncated expectations
using (\ref{eq:tr4tomom}).  Since $\FT{V}$ is real, 
all terms involving second order truncated correlation functions produce terms which are purely imaginary
and, hence, they do not contribute to the derivative of the Wigner function.  Therefore,
\begin{align}\label{eq:spinlessWevol3}
 & \partial_t  W(k,t) 
 = \re\Bigl[\ci \lambda \int_{(\Lambda^*)^4} \! \rmd k_1 \rmd k_2\rmd k_3\rmd k_4\,
  \left(\FT{V}(k_2+k_3)-\FT{V}(k_1+k_3)\right)
   \nonumber \\ & \qquad  \times 
  \delta_\Lambda( k-k_1-k_2-k_3) \rho^T[a(k_1,1,t), a(k_2,1,t), a(k_3,-1,t), a(k_4,-1,t)]\Bigr]
 \,.
\end{align}

Computation of derivatives of higher order truncated correlation functions would be simplified 
by introducing the associated Wick polynomials, as was observed in \cite{LM16} for commuting fields.
However, it is still possible to work out the necessary combinatorics and cancellations by hand for the fourth 
order terms which are needed to compute the collision operator of kinetic theory. 
Namely, after a somewhat lengthy computation employing the symmetry of the function $\FT{V}$, one finds that
\begin{align}\label{eq:rhot4evol}
 & \partial_t\left(\rme^{-\ci t(\omega_1+\omega_2-\omega_3-\omega_4)}\rho^T[a(k_1,1,t),a(k_2,1,t),a(k_3,-1,t), a(k_4,-1,t)] \right)
 \nonumber \\&\quad =
  \ci \lambda \rme^{-\ci t(\omega_1+\omega_2-\omega_3-\omega_4)} \delta_\Lambda(k_1+k_2+k_3+k_4)
   \left(\FT{V}(k_2+k_3)-\FT{V}(k_1+k_3)\right)
  \nonumber \\&\qquad  \times
  \Bigl[ \tilde{W}(k_2) W(-k_3) W(-k_4) - W(k_1)W(-k_3) W(-k_4) 
  \nonumber \\&\qquad\quad 
  + W(k_1)W(k_2) W(-k_4) - W(k_1)W(k_2) \tilde{W}(-k_3) \Bigr]
  \nonumber \\&\qquad
  + (\text{higher order truncated functions})\,,
\end{align}
where we have introduced the shorthand notations $\omega_i := \omega(k_i)$, $\tilde{W}=1-W$,
and each $W$ and $\tilde{W}$ factor is evaluated at $t$.

We then integrate the above time-derivatives from $0$ to $t$.
The terms involving higher order truncated functions (4:th and 6:th in (\ref{eq:rhot4evol})), as well as the substitution term involving
the 4:th order truncated correlation at time $0$,
are expected to contribute only terms which are subleading in $\lambda$ at the kinetic time scales $t\propto \lambda^{-2}$, due to the ``integrals'' over the oscillatory phase factors.  The remaining terms yield the approximation 
\begin{align}\label{eq:spinlessWevol4}
 &  W(k,t) - W(k,0) \approx  \int_0^t\!\rmd t' \int_0^{t'}\!\rmd s
 \re\Bigl\{-\lambda^2 \int_{(\Lambda^*)^4} \! \rmd k_1 \rmd k_2\rmd k_3\rmd k_4\,
 \rme^{\ci (t'-s)(\omega_1+\omega_2-\omega_3-\omega_4)}
   \nonumber \\ & \qquad  \times 
  \left(\FT{V}(k_2+k_3)-\FT{V}(k_1+k_3)\right)^2
  \delta_\Lambda( k-k_1-k_2-k_3)  \delta_\Lambda(k_1+k_2+k_3+k_4)
   \nonumber \\ & \qquad  \times   
  \Bigl[ \tilde{W}(k_2) W(-k_3) W(-k_4) - W(k_1)W(-k_3) W(-k_4) 
  \nonumber \\&\qquad\quad 
  + W(k_1)W(k_2) W(-k_4) - W(k_1)W(k_2) \tilde{W}(-k_3) \Bigr] \Bigr\}
 \,,
\end{align}
where
each $W$ and $\tilde{W}$ factor is evaluated at $s$.  Inside 
the integrand $-k_4=k$.  Hence, integration over $k_4$ is straightforward
and 
swapping the sign of $k_3$, the order of time-integrals, and denoting 
$W_i:=W(k_i,s)$ and $\tilde{W}_i := 1- W_i$, we arrive at the approximation
\begin{align}\label{eq:preBoltzmannspinless}
 &  W(k_0,t) - W(k_0,0) \approx \lambda^2 \int_0^t\!\rmd s
 \int_{(\Lambda^*)^3} \! \rmd k_1 \rmd k_2\rmd k_3\,
 \re \int_0^{t-s}\!\rmd r\, \rme^{\ci r(\omega_1+\omega_2-\omega_3-\omega_0)}
   \nonumber \\ & \qquad  \times 
  \left(\FT{V}(k_2-k_3)-\FT{V}(k_1-k_3)\right)^2
  \delta_\Lambda( k_0-k_1-k_2+k_3) 
   \nonumber \\ & \qquad  \times   
  \Bigl[ -\tilde{W}_2 W_3 W_0+ W_1 W_3 W_0 - W_1 W_2 W_0 + W_1 W_2  \tilde{W}_3\Bigr]
 \,,
\end{align}

The real part of the remaining oscillatory time-integral formally convergences to 
$\pi \delta(\omega_0-\omega_3-\omega_1-\omega_2)$ as $t\to \infty$.  
In fact, the $\delta$-function approximation should
only be used after the thermodynamic limit $L\to \infty$ has been taken;
for a finite lattice, also values for which 
$\omega_1+\omega_2-\omega_3-\omega_0$ is not exactly zero but 
close enough to zero (e.g., $o(L^{-2})$), will contribute to the collision term.
Assuming that the thermodynamic limit of the function $W$ exists
and using the same notation for the limit, 
we obtain
\begin{align}\label{eq:fermBNeqn}
 &  W(k_0,t) - W(k_0,0) \approx \int_0^t\!\rmd s\, \CFBN[W(\cdot,s)](k_0)
 \,,
\end{align}
where a relabelling $k_1\leftrightarrow k_3$  yields the following more standard form of a fermionic
Boltzmann--Nordheim collision operator
\begin{align}\label{eq:BNcollision}
 & \CFBN[W](k_0) :=
  \pi \lambda^2 
 \int_{(\T^d)^3} \! \rmd k_1 \rmd k_2\rmd k_3\,
 \delta(\omega_0+\omega_1-\omega_2-\omega_3)
   \nonumber \\ & \qquad  \times 
  \left(\FT{V}(k_1-k_2)-\FT{V}(k_1-k_3)\right)^2
  \delta_{\T^d}( k_0+k_1-k_2-k_3) 
   \nonumber \\ & \qquad  \times   
  \Bigl[ \tilde{W}_1  W_2 W_3  - W_0  W_2 W_3-  W_0  W_1 \tilde{W}_2 +  W_0 W_1 W_3  \Bigr]
 \,.
\end{align}

The kinetic equation obtained by replacing the approximation sign in (\ref{eq:fermBNeqn}) 
by an equals sign is called the (spatially homogeneous) fermionic Boltzmann--Nordheim equation. 
The term in square brackets in (\ref{eq:BNcollision}) is then 
usually written in a more symmetric form as
\[
  \tilde{W}_0\tilde{W}_1 W_2 W_3 - W_0 W_1\tilde{W}_2\tilde{W}_3  \,.
\]
However, it should be noted that, since the highest order terms indeed cancel, the 
collision operator has a nonlinearity of third order, not of fourth order.

The above lattice kinetic theories have two conserved quantities,
$\int \rmd k \, \omega(k) W(k,t)$ related to energy and  $\int \rmd k \, W(k,t)$ related to particle density.
The mathematical properties of their solutions have mainly been studied 
in the continuum case for which instead of the lattice wave number $k\in \T^d$
one uses the particle velocity $v\in\R^d$ and the dispersion relation is $\omega(v)= v^2$ in the nonrelativistic case.
For the existence and uniqueness of solutions in the continuum case, we refer to \cite{dolb94,EMV03},
while the corresponding issues for a lattice model will be discussed in the next section, based on \cite{LMS12}.

\subsection{Kinetic theory of the spatially homogeneous Hubbard model}

We next repeat the above computations for the Hubbard model which has a simple onsite potential but includes 
spin-interactions.  By (\ref{eq:Wevol}),
\begin{align}\label{eq:HubbardWevol}
 & \partial_t  W_{\sigma'\sigma}(k,t) 
 =  \int_{\Lambda^*} \!\rmd k'\, 
 \rho[\partial_t a(k,\sigma',1,t) a(k',\sigma,-1,t)]+ (\text{h.c.})
 \,,
\end{align}
where ``h.c.'' denotes a Hermitian conjugate with respect to the spin degrees of freedom.
Employing (\ref{eq:atauHubbard}) we find
\begin{align}\label{eq:Hubbardder}
&   \int_{\Lambda^*} \!\rmd k'\, 
 \rho[\partial_t a(k,\sigma',1,t) a(k',\sigma,-1,t)]
 = \ci \omega(k) \int_{\Lambda^*} \!\rmd k'\,
  \rho[a(k,\sigma',1,t) a(k',\sigma,-1,t)]
   \nonumber \\ & \qquad +
  \ci \lambda \int_{(\Lambda^*)^4} \! \rmd k_1 \rmd k_2\rmd k_3\rmd k_4\,
  \delta_\Lambda( k-k_1-k_2-k_3)
   \nonumber \\ & \qquad\quad  \times 
   \rho[a(k_1,\sigma',1,t) a(k_2,-\sigma',1,t) a(k_3,-\sigma',-1,t) a(k_4,\sigma,-1,t)]\,.
\end{align}

The first term on the right is antisymmetric with respect to the Hermitian conjugate, and hence does not contribute to the time derivative of $W$.  
We represent the remaining expectation in terms of truncated expectations
using (\ref{eq:tr4tomom}).  In contrast to the spinless case, the second order terms need no longer cancel:
explicitly, they contribute to (\ref{eq:Hubbardder}) the term
\begin{align}\label{eq:Hubbardsecond}
& \ci \lambda \int_{\Lambda^*} \! \rmd k' \left(W_{\sigma'\sigma}(k) W_{-\sigma',-\sigma'}(k')
-W_{-\sigma',\sigma}(k) W_{\sigma',-\sigma'}(k')\right)\,.
\end{align}
It depends on the expectation
\begin{align}\label{eq:defSigma}
 \Sigma_{\sigma'\sigma} := \int_{\Lambda^*} \! \rmd k' \, W_{\sigma'\sigma}(k')
 = \rho[a^*(0,\sigma') a(0,\sigma)] = \frac{1}{|\Lambda|} \sum_{x\in \Lambda}
  \rho[a^*(x,\sigma') a(x,\sigma)]\,,
\end{align}
i.e., on the spin correlation matrix.  These
expectations are conserved by the time evolution of the Hubbard model, 
and hence the matrix $\Sigma_{\sigma'\sigma}$ is time-independent.
Therefore, the dominant term in the time-derivative (\ref{eq:HubbardWevol})
is given by 
\begin{align}\label{eq:RWcommHubbard}
& \ci \lambda \left(W_{\sigma'\sigma}(k) \Sigma_{-\sigma',-\sigma'}
-W_{-\sigma',\sigma}(k) \Sigma_{\sigma',-\sigma'}
-W_{\sigma'\sigma}(k) \Sigma_{-\sigma,-\sigma}
+W_{\sigma',-\sigma}(k) \Sigma_{-\sigma,\sigma}\right)
\,,
\end{align}
which is most conveniently written as the $(\sigma',\sigma)$ -component
of the commutator
\[
 -\ci \lambda [ \Sigma, W(k,t)]\,.
\]

New terms arise also in the computation of the second order term in $\lambda$.  
The computations are in principle completely analogous to those in the previous subsection
but one has to carefully consider the propagation of the spin variable.
After taking the thermodynamic limit $L\to \infty$ and neglecting terms which are expected to 
be higher order in $\lambda$, new features compared to the spinless case arise.
Most importantly, since one takes a Hermitian, not complex, conjugate of (\ref{eq:Hubbardder}), the imaginary
part of the oscillatory time-integral also contributes in the evolution equation.  
In other words, one needs to use here the formal identification
\[
 \int_0^{\infty}\!\rmd r\, \rme^{\ci r \omega}
 = \pi \delta(\omega) + \ci \, \text{P.V.} \frac{1}{\omega}\,,
\]
where ``P.V.'' denotes a Cauchy principal value when integrating over the real variable $\omega$.
The terms arising from the imaginary part do not resemble usual collision integrals.  
Instead, they combine into conservative Vlasov-type
terms, similarly to what occurred above for the lowest order contribution.

The final evolution equation is most conveniently written as an evolution equation for the 
Hermitian $2\times 2$ -matrix $W(k,t)$, $k\in \T^d$.  It reads
\begin{align}\label{eq:HubbardBE}
 \partial_t W(k,t) = \CHubb[W(\cdot,t)](k) - \ci \left[ \Heff[W(\cdot,t)](k),W(k,t)\right]  \,,
\end{align}
where the collision operator may be written as
\begin{align}\label{eq:defHBcoll}
 & \CHubb[W](k_0) :=  \lambda^2 \pi\int_{(\T^{d})^3}\!\! \rmd k_1 \rmd k_2 \rmd k_3\,
\delta(k_0+k_1-k_2-k_3) \delta(\omega_0+\omega_1-\omega_2-\omega_3)  \nonumber \\ & \quad
\times\Bigl( 
\tilde{W}_0 W_2 J[\tilde{W}_1 W_3]
+ J[ W_3 \tilde{W}_1] W_2 \tilde{W}_0 
 -W_0\tilde{W}_2 J[W_1\tilde{W}_3]
- J[\tilde{W}_3 W_1]\tilde{W}_2  W_0\Bigr)
\end{align}
using the matrix operation
$J[A]:= 1\, \tr A-A\in \C^{2\times 2}$.  The ``effective Hamiltonian'' in the matrix commutator term 
is given by 
\begin{align}\label{eq:defHeff}
 &  \Heff [W](k_0) := \lambda \Sigma + \lambda^2 \text{P.V.} \int_{(\T^{d})^3} \rmd k_1 \rmd k_2 \rmd k_3
\delta(k_0+k_1-k_2-k_3)\nonumber \\ & \quad
\times  \frac{1}{\omega_0+\omega_1-\omega_2-\omega_3} \left( \tilde{W}_2 J[W_1\tilde{W}_3] 
+ W_2 J[\tilde{W}_1 W_3]\right) \,.
\end{align}

Also the Hubbard--Boltzmann equation (\ref{eq:HubbardBE}) can be derived using 
direct perturbation expansions and their graph representations, as has been done in \cite{flms12}
for more general spin-interaction potentials and with a slightly different splitting between 
the terms in $H_0$ and $V$ operators.  Neither of these derivations provides rigorous estimates 
of how accurately 
the solutions to the Hubbard--Boltzmann equation describe the original fermionic reduced density matrices.
The principal value integral, in particular, is somewhat troublesome from a mathematical point of view.

The precise mathematical meaning of the terms appearing in 
the Hubbard--Boltzmann equation (\ref{eq:HubbardBE}), as well as the existence and uniqueness of its solutions for 
physically relevant initial data, have been studied in \cite{LMS12}.
It is shown there that for the nearest neighbour Hubbard model with a sufficiently high dimension, $d\ge 3$,
any Lebesgue measurable initial data $W_0(k)$ satisfying the matrix constraint $0\le W_0(k)\le 1$
allows a global solution to (\ref{eq:HubbardBE}) which is also unique among solutions satisfying the 
constraint $0\le W(k,t)\le 1$.  (The constraint is physically related to the Pauli exclusion principle and 
it can be checked to follow from the earlier mentioned properties of the fermionic creation and annihilation operators.)
This solution is also proven to conserve energy and total spin.  More precisely, the
real observable 
$\int\!\rmd k\, \omega(k) \tr W (k,t)$ and the matrix observable $\int\!\rmd k\, W(k,t)$
are constants along the solutions.  Together these properties show that the approximations
leading to the Hubbard--Boltzmann equation are consistent, and the resulting kinetic equation should
have range of validity similar to the more standard kinetic theories such as the Boltzmann--Nordheim equation derived earlier.

\section{Thermalization in spatially homogeneous kinetic theory}

For ergodic systems, time averages of observables 
will converge to ensemble averages when the averaging period is taken to infinity.
In fact, the ensembles covered by such limits could be identified with thermal equilibrium states of the system.
However, for system with local conservation laws the approach to global equilibrium
typically takes a very long time, often diverging when the system size is increased:
for instance, for systems with normal heat conductivity heat 
relaxation occurs diffusively and thus involves time-scales of order $L^2$ for systems of spatial diameter 
$L$.

For physical transport phenomena one is interested in the state of the system
at \defem{mesoscopic} timescales, i.e., times which are long in microscopic units but short on the 
macroscopic scale.  If the system has only short range interactions, 
even though its state could not yet be well approximated by the global equilibrium state, often  
time-averages of observables local to a point in space can be ever better approximated by one of the equilibrium states.
This allows describing the evolution of the state of the system by first parametrizing its equilibrium states and then inspecting the evolution
of these parameters.  A common example would be introduction of space-time dependent temperature function
related to the temperature parameter 
of the canonical Gibbs state for those systems where total energy is conserved by the evolution.

Systems, which have the above local approximation property, are said to be in \defem{local thermal equilibrium},
and \defem{thermalization} refers to the approach to 
one of the local thermal equilibrium states from the given initial state.  The \defem{thermalization time}, i.e., the time 
it takes for local thermal equilibrium states to become good approximations, is typically mesoscopic, not macroscopic.

In fact, kinetic theory provides a method of estimating the thermalization process and times.
We focus here on thermalization of spatially homogeneous states.  This simplifies the analysis since the slow processes
associated with spatial relaxation of the equilibrium parameters are then absent.   
As explained below, kinetic theory indicates that the Wigner function relaxes to stationary 
states labelled by a few parameters and hence one would expect local equilibrium or quasi-equilibrium to 
be reached already at kinetic timescales proportional to $\lambda^{-2}$.  The key to 
these properties is finding an entropy functional
satisfying an H-theorem for the appropriate kinetic evolution.  The 
vanishing of entropy production restricts the functional form of stationary solutions and allows their explicit parametrisation.

\subsection{Thermalization without spin-interactions}

The entropy functional associated with the spatially homogeneous fermionic Boltzmann--Nordheim equation,
\[
 \partial_t W(k,t) = \CFBN[W(\cdot,t)](k)\,,
\]
where the collision operator is defined in (\ref{eq:BNcollision}), is given by 
\begin{align}
 S[W] := -\int_{\mathbb{T}^d} \rmd k \left( W(k)\log W(k)+
\widetilde{W}(k)\log\widetilde{W}(k)\right)\,.
\end{align}
Computing the time-derivative, one obtains
\[
 \frac{\rmd}{\rmd t}S[W(t)]=\sigma[W(t)]\,,
\]
where the \defem{entropy production functional} is 
\begin{align}
 &\sigma[W]= \pi \int_{(\mathbb{T}^d)^4} \rmd k_1 \rmd k_2
\rmd k_3 \rmd k_4 \delta(k_1+k_2-k_3-k_4)\delta(\omega_1
+\omega_2-\omega_3-\omega_4)\nonumber\\
& \qquad \times \left(\FT{V}(k_2-k_3)- \FT{V}(k_2-k_4)\right)^2
G(\tilde{W}_1\tilde{W}_2 W_3 W_4,W_1 W_2\tilde{W}_3\tilde{W}_4)\,,
\end{align}
with $G(x,y)=(x-y)\ln (x/y)$.   Since $\sigma[W]\ge 0$ for physical Wigner functions with $W,\tilde{W}\ge 0$, this proves that $S$ satisfies an analogue of the 
\defem{H-theorem} of classical rarefied gas Boltzmann equation. 

In particular, any stationary solution to the kinetic equation needs to satisfy $\sigma[\Weql]=0$.
For sufficiently non-degenerate $\FT{V}$ and $\omega$, the only regular solutions to this equation
are given by the two-parameter family
\begin{align}\label{eq:BNWeql}
 \Weql_{\beta,\mu}(k)=\big(\mathrm{e}^{\beta(\omega(k)-\mu)}+ 1\big)^{-1}\,, 
\end{align}
where the values of the parameters $\beta,\mu\in\R$ could also be fixed by giving the values for the 
conserved energy and particle density observables.  These Wigner functions can also be obtained by considering the one-particle
reduced density matrix of the standard grand canonical Fermi--Dirac states after setting $\lambda=0$, cf.\ \cite[Proposition 5.2.23]{bratteli:ope2}.  
These states are gauge invariant and quasifree and thus the Wigner function determines all other reduced density matrices.

It is clear that $\FT{V}(k)$ cannot be a constant since then $\CFBN[W]=0$, 
but otherwise the function $\FT{V}$ can be fairly arbitrary for this result to hold; one merely needs that the 
difference $\FT{V}(k_2-k_3)- \FT{V}(k_2-k_4)$ is nonzero almost everywhere on the manifold defined by the two $\delta$-constraints.
The conditions on the dispersion relation $\omega$ are more intricate but in two and higher dimensions quite generally the above solutions should be the 
only stationary ones, see \cite[Appendix B.1]{ls09} and \cite{spohn06} for detailed conditions and more discussion on the topic.

In case $\FT{V}$ and $\omega$ are such that the only stationary solutions are given by (\ref{eq:BNWeql}), 
one expects that for any regular initial data the solution of the fermionic Boltzmann--Nordheim equation converges as $t\to \infty$ to the unique
function $\Weql_{\beta,\mu}$ where $\beta,\mu\in\R$ are determined by the initial energy and particle number.
Unlike for the corresponding bosonic equation, the solutions cannot diverge since they satisfy $0\le W\le 1$ at all times. Thus 
the space of regular stationary solutions should suffice to cover all asymptotic limits of the solutions.  
The convergence to a regular stationary solution has been proven for certain continuum models and initial data in \cite{Lu2003}.

The above results suggest that thermalization timescale for weakly interacting spinless lattice fermions
is in great generality given by the kinetic timescale, $t\propto \lambda^{-2}$.  It is also consistent with the hypotheses
that, apart from special degenerate interactions, the only equilibrium parameters are related to the conservation of energy and particle number.
More precisely, one can use $\beta$ and $\mu$ of the standard grand canonical Fermi--Dirac states on the fermionic Fock space as parameters.

\subsection{Thermalization in the Hubbard model}

The spin-structure of the Hubbard--Boltzmann equation (\ref{eq:HubbardBE}) leads to some new phenomena compared to the above spinless Boltzmann--Nordheim
case.  The entropy functional needs to be generalised to 
\begin{align}
 S[W] := -\int\!\rmd k\left(\tr W \ln W + \tr \tilde{W}\ln \tilde{W}\right)\,,
\end{align}
where $W$ is a $2\times 2$ Hermitian matrix.  Computing its derivative requires some effort, yielding
\[
 \frac{\rmd}{\rmd t}S[W(t)]=\sigma[W(t)]\,,
\]
where the entropy production functional is again positive, $\sigma[W]\ge 0$.  To write down the entropy production,
let us first diagonalize the matrices $W(k)$, yielding an eigensystem $(\lambda_a(k),\psi_a(k))$, $a=1,2$, for each $k\in \T^d$.
Then 
\begin{align*}
& \sigma[W](k_1) := \frac{\pi}{4} \int\! \rmd^4 k\, \delta(k_1+k_2-k_3-k_4)\delta(\omega_1+\omega_2-\omega_3-\omega_4) \sum_{a\in \{1,2\}^4}
\\ & \quad \times 
\left( \tilde{\lambda}_1 \tilde{\lambda}_2 \lambda_3 \lambda_4- \lambda_1 \lambda_2\tilde{\lambda}_3 \tilde{\lambda}_4\right)
\ln \frac{\tilde{\lambda}_1 \tilde{\lambda}_2 \lambda_3 \lambda_4}{\lambda_1 \lambda_2\tilde{\lambda}_3 \tilde{\lambda}_4}
\left| \braket{\psi_1}{\psi_3}\braket{\psi_2}{\psi_4}-\braket{\psi_1}{\psi_4}\braket{\psi_2}{\psi_3}\right|^2\,,
\end{align*}
where $\psi_i := \psi_{a_i}(k_i)$, $\lambda_i := \lambda_{a_i}(k_i)$ and $\tilde{\lambda} := 1-\lambda$.

The solution of the condition $\sigma[W]=0$ is no longer quite as straightforward as before, and one has to consider
a few degenerate cases separately.  However, if $d\ge 2$, the non-degeneracy conditions mentioned earlier are satisfied for the nearest neighbour 
interaction of the Hubbard model, and thus the analysis of the two $\delta$-constraints is simplified.
As derived in \cite{FMS12a}, then one of the following possibilities needs to be realized by physical stationary solutions $\Weql(k)$ 
which are Hermitian matrices satisfying $0\le W(k)\le 1$ for every $k\in \T^d$.  
First, choose a spin-basis such that the total spin-correlation matrix $\Sigma$ is diagonal.  Then one of the following cases holds:
\begin{enumerate}
 \item There are grand canonical parameters $\beta,\mu_+, \mu_-$, fixed by the diagonal matrix $\Sigma$ and the energy,
 such that 
 \begin{align}\label{eq:HubbardWeql}
 \Weql(k) = \begin{pmatrix} g_+(k) & 0 \\ 0 & g_-(k) \end{pmatrix}\,, 
 \end{align}
  where $g_{\pm}(k) := (1+\rme^{\beta (\omega(k)-\mu_{\pm})})^{-1}$ are standard Fermi--Dirac distributions.
 \item One of the bands is empty and the other is arbitrary: there is a function $f(k)$ with $0\le f(k)\le 1$ and 
 $\sigma\in \set{\pm 1}$
 such that $W_{\sigma\sigma}(k)=f(k)$ and all other elements of $W(k)$ are zero.
 \item One of the bands is full and the other is arbitrary: there is a function $f(k)$ with $0\le f(k)\le 1$ and 
 $\sigma\in \set{\pm 1}$
 such that $W_{\sigma\sigma}(k)=f(k)$, $W_{-\sigma,-\sigma}(k)=1$, and all off-diagonal elements of $W(k)$ are zero. 
\end{enumerate}

These solutions are expected to behave differently when occurring as asymptotic stationary states in the Hubbard model.
If the initial data is such that both bands are partially filled, i.e., if one can find $\beta,\mu_+, \mu_-$
and a unitary matrix $U$ such that the function $\Weql$ in (\ref{eq:HubbardWeql}) 
satisfies $\int \rmd k\, U^* W(k,t) U = \int \rmd k\, \Weql(k)$
and $\int \rmd k\, \omega(k) \tr W(k,t) = \int \rmd k\, \omega(k) \tr \Weql(k)$ initially, and hence for all $t$, then one expects 
$ W(k,t) \to U \Weql(k) U^*$ as $t\to \infty$.

However, if one of the bands is either empty or full initially, then no thermalization can be expected.
In fact, this property is not only an artefact of the kinetic theory but it can  also be realised
in the original Hubbard model.  Consider an initial wave vector for which there are no particles with $-$ -spin.
Then the pair-interaction $V$ acting on the vector produces zero and, since the free Hamiltonian does not mix the two bands,
one can check that Hubbard model evolution equations are satisfied by the solution of the free evolution
generated by $H_0$.  The free semigroup 
leaves for instance all quasifree states invariant and one can choose the Wigner function of the $+$ -component arbitrarily.

The above situation is radically changed if $d=1$.  This case is known to be integrable, 
see \cite{essler_etal_2005} for a review of the one-dimensional Hubbard model, and the large number of conserved quantities is
reflected also in the kinetic evolution.  As shown in \cite{FMS12a}, in this case one may take 
in the stationary solutions in item 1 above instead of the standard
Fermi--Dirac distributions $g_\pm$ any functions which are of the form $(1+\rme^{\beta (f(k)-\mu_{\pm})})^{-1}$
for some real periodic function $f$ which satisfies the antisymmetry condition $f(\frac{1}{2}-k)=-f(k)$.  Hence, one needs infinitely many 
parameters to describe the stationary solutions.  The various scenarios for the convergence towards a steady state are explored numerically in \cite{FMS12a}. There it is also observed that adding a next-to-nearest neighbour term to the free evolution appears to lift the degeneracy, leaving only the standard
Fermi--Dirac distributions as possible limits, similarly to what was stated above for the cases with $d\ge 2$.

\section{Concluding Remarks}

Reliable study of large scale evolution of a system of weakly interacting fermions is a challenge both to
numerical simulations and to theoretical analysis.  We advocate here 
using kinetic equations not only to reproduce standard folklore results, such as convergence towards Fermi--Dirac distribution,
but as a tool for \defem{systematic} study of the approach to equilibrium and thermalization in these systems.
Even lacking complete mathematical control over the accuracy and applicability of the kinetic approximation, 
analysis of kinetic equations can provide testable predictions and reveal possible sources of ``anomalies'' and 
other degeneracies.  For instance, the role of the dispersion relation and dimensionality in the Hubbard 
model revealed in the above references encourages such studies in other models.

The almost unreasonable usefulness of kinetic theory begs for better understanding of its underpinnings,
in particular, of what is the most accurate connection between the microsopic evolution 
and the kinetic theory and
what are the most appropriate kinetic equations for this purpose.  
These questions lie in the realm of mathematically rigorous study of scaling limits producing
observables which exactly follow some kinetic equation.  However, ultimately the goal should be in also extracting 
practical information about the error in such approximations and how well the approximations
extend beyond their apparent regions of applicability, as dictated by the convergence of the scaling limits.

For instance, finding answers to the following open questions could benefit from mathematically rigorous approaches:
\begin{enumerate}
 \item For which initial data does the corresponding solution to the kinetic equation converge towards the stationary solution determined by the values of the conserved quantities?  Could one estimate the rate of convergence?
 \item How would the kinetic equations and their solutions change for general spin-interactions, including also interactions with external magnetic fields?
 \item If the initial state of the system is not spatially homogeneous, when does its evolution follow an inhomogeneous Boltzmann equation?  Are there ways of improving the accuracy of the model, for instance, by including a Vlasov-Poisson-type correction?  
 \item Could one improve the accuracy of the kinetic equation by ``renormalizing'' the microscopic observables?  How much?
\end{enumerate}

\subsection*{Acknowledgements}

I am most grateful to Herbert Spohn for our collaboration and many discussions about validity and properties of kinetic theory.
Most of the results here are based on his works and on our joint collaborations.
The related research has been made possible by support from the Academy of Finland and also partially supported by
the French Ministry of Education through the grant ANR (EDNHS).

\end{document}